\documentclass[12pt]{iopart}
\usepackage{graphicx}
\usepackage{iopams}

\def\be{\begin{equation}}
\def\ee{\end{equation}}
\def\ber{\begin{eqnarray}}
\def\eer{\end{eqnarray}}
\def\bern{\begin{eqnarray*}}
\def\eern{\end{eqnarray*}}

\def\rv{\mathbf{r}}

\def\jv{\mathbf{j}}
\def\Gv{\mathbf{G}}

\def\pv{\mathbf{p}}
\def\qv{\mathbf{q}}

\def\Ev{\mathbf{E}}

\def\0v{\mathbf{0}}
\def\1v{\mathbf{1}}
\def\2v{\mathbf{2}}
\def\3v{\mathbf{3}}

\begin{document}

\title[Electronic excitations in quasi-2D crystals]{Electronic excitations in 
quasi-2D crystals: What theoretical quantities are relevant to experiment?}

\author{V. U. Nazarov}
\address{Research Center for Applied Sciences,  Academia Sinica, Taipei 11529, Taiwan}

\maketitle

\begin{abstract} 
The {\it ab initio} theory of electronic excitations in atomically thin  [quasi-two-dimensional (Q2D)] crystals  presents
extra challenges in comparison to both the bulk and purely 2D cases. 
We argue that the conventionally used  energy-loss function
$- {\rm Im}\,\frac{1}{\epsilon(\qv,\omega)}$ 
(where $\epsilon$, $\qv$, and $\omega$ are the dielectric function, the momentum, and the energy transfer,
respectively) is not, generally speaking, the suitable quantity for the interpretation
of the electron-energy loss spectroscopy (EELS) in the Q2D case, and we construct
different functions pertinent to the EELS experiments on Q2D crystals.
Secondly, we emphasize  the  importance and develop a convenient procedure of the  
elimination of the spurious inter-layer interaction
inherent to the use of the 3D super-cell  method for the calculation of excitations in Q2D crystals.
Thirdly, we resolve the existing controversy in the interpretation of the so-called $\pi$ and $\pi+\sigma$ 
excitations  in monolayer graphene by 
demonstrating that both dispersive collective excitations (plasmons) and non-dispersive 
single-particle (inter-band) transitions fall in the same energy ranges, where they strongly influence each other.
\end{abstract}

\section{Introduction}
\label{Intr}

Quasi two-dimensional (Q2D) crystals (the systems periodic and infinite  in two dimensions while of  atomic-size thickness) are presently attracting much of attention because of their potential in future
nano-technologies. In particular, the  electronic excitations in Q2D crystals, such as electron-hole pairs generation, plasmons, excitons, {\it etc.},
are of great importance, since they determine the response to external actions, and, ultimately, shape the materials' functionality (see, e.g., \cite{Politano-14} for a recent review). At the same time, the theoretical approaches as well as the computational methods to deal with the  excitations in Q2D systems
become much more involved compared to both the three-dimensional (3D) and purely 2D cases, and  are still far from finally established.

Indeed, in both purely 3D and purely 2D cases (the latter referring to model systems of zero thickness),
one usually introduces the dielectric function $\varepsilon(\qv,\omega)$, where $\qv$ is the wave-vector of
the corresponding dimensionality, and $\omega$ is the frequency.  The definition of $\varepsilon$ reads
\footnote{In this paper we are concerned with the longitudinal fields.}
\begin{equation}
\hspace{2 cm}
\phi^{ext}(\qv,\omega) = \varepsilon(\qv,\omega) \phi^{tot}(\qv,\omega),
\label{def}
\end{equation}
where $\phi^{ext}$ and $\phi^{tot}$ are the externally applied  and the total (external plus induced)
scalar potentials of the electric field, respectively. The definition (\ref{def}) is not, however, straightforwardly
transferable to the Q2D case,
which can be easily appreciated by considering that $\phi^{tot}$ depends on the $z$ (perpendicular to the
layer) coordinate, even when $\phi^{ext}$ is uniform in $z$ direction. 
Therefore, we need to redefine the dielectric function (energy-loss function, conductivity, {\it etc.})
of a Q2D crystal in accordance to the system's structure. Importantly, the new definitions must be consistent
with the quantities measured experimentally, the violation of which requirement having caused much of confusion 
in recent literature.

Secondly, the widest used at this time method of dealing with Q2D systems numerically is
the super-cell geometry approach. This is to artificially replicate the system
periodically in $z$ direction, choosing the distance $d$
between the layers large enough to prevent the inter-layer interactions.
The  resulting fictitious
system is 3D-periodic, and well developed 3D solid-state methods as well as the existing 3D software can be used
to efficiently handle it.
In the ground-state calculations, the interpretation of the super-cell results is straightforward:
Provided that the inter-layer distance  is such that
the corresponding wave-functions do not overlap, properties of an isolated single layer coincide
with those of a single super-cell. 

However, if we are concerned with the  dynamic response
to electric field, 
the situation changes. In the case of a Q2D crystal, 
the electric field decays into vacuum as $\sim e^{- q |z|}$. Clearly, whatever large is the separation $d$ between
the layers, at sufficiently small $q$ the inter-layer interaction persists,
and results of the super-cell calculation cannot be literally transferred to a single layer.
A simple method of taking $d$ very large does not work, because  small $q$-s are often of special interest,
and the computational load grows rapidly with the increase of $d$.
Therefore, we need to establish relations between the response functions of the fictitious
3D array system and those of the single layer of interest. The paramount requirement to these relations ({\em test for sanity}) is that they should produce the same response functions of a single layer
from the differing 3D response functions calculated with different separations $d$.

In this paper, in the framework of the time-dependent density-functional theory (TDDFT), 
we construct a scheme satisfying the above requirements.
As its practical application, we calculate and perform a detailed analysis of the excitation spectrum of a single-layer graphene in the 1-30 eV energy range.
Resolving the recent controversy 
in the interpretation of the $\pi$ and $\pi+\sigma$ peaks as plasmons \cite{Gass-08,Eberlein-08,Liu-08,Liou-15}
or single-particle inter-band transitions \cite{Nelson-14},
we show that, while the problem is far more complicated than thought before, there {\em  do exist} prominent $\pi$ and $\pi+\sigma$ plasmons in graphene, although they overlap with the inter-band transitions. 
The isolation of the both kinds of excitations becomes possible by following the wave-vector dependence of spectra
(spatial dispersion).
We also show that much of the previous confusion was due to using unsuitable theoretical quantities
for comparison with experiment, as well as 
to the uncritical use of the results of the super-cell calculations in application to  Q2D systems.

This paper is organized as follows. In Sec.~\ref{LR} we re-examine and revise the definitions of
the response-functions (dielectric function, energy-loss function, conductivity)
pertinent to measurements on Q2D crystals. Sec.~\ref{Extr} is devoted to the methods of obtaining
the response-functions of Q2D crystals from the 3D super-cell calculations.
In Sec.~\ref{RES} we present results of calculations performed for  graphene as an important example
of a Q2D crystal, discuss the differences arising from the use of our theory, and compare with experiment.
Section~\ref{Concl} contains  conclusions, and in the Appendix we give the derivation
of some formulas presented in Sec.~\ref{Extr}.
We use the atomic units ($e^2=\hbar=m_e=1$) unless explicitly specified otherwise.

\section{Linear response of quasi-2D crystals: Definition of  quantities}
\label{LR}

Quasi-2D crystals are neither periodic three-dimensionally nor strictly
two-dimensional within the mathematical plane: They rather have a finite atomic or nano-scale thickness.
Accordingly, one has to exercise caution in the choice of quantities 
(dielectric function, conductivity, density-response function, {\it etc.})
which characterize the dielectric response,
giving them rigorous, suitable for Q2D systems, meaning.
The safe way to do this is to start from the general 3D case without assuming the periodicity (translational invariance) in any direction.
The density-response function $\chi(\rv,\rv',\omega)$ is defined in the real-space representation as
\begin{equation}
\rho(\rv,\omega)= \int \chi(\rv,\rv',\omega) \phi^{ext}(\rv',\omega) d\rv',
\label{chi3r}
\end{equation}
where $\phi^{ext}$ is the scalar potential of the externally applied electric field
and $\rho$ is the charge-density induced in the system
in response to $\phi^{ext}$. 
The inverse generalized dielectric function $\varepsilon^{-1}(\rv,\rv',\omega)$ is defined as
\begin{equation}
\phi(\rv,\omega) = \int \varepsilon^{-1}(\rv,\rv',\omega) \phi^{ext}(\rv',\omega) d\rv',
\end{equation}
where $\phi$ is the total (external plus induced) potential,
and it is expressed through $\chi$ as
\begin{equation}
\varepsilon^{-1}(\rv,\rv',\omega)=\delta(\rv-\rv')+\int \frac{\chi(\rv'',\rv',\omega)}{|\rv-\rv''|}  d\rv''.
\label{ecr}
\end{equation}
By Fourier-transform, Eqs.~(\ref{chi3r})-(\ref{ecr}) can be equivalently written in the reciprocal-space representation,
of which we will need only
\begin{equation}
\varepsilon^{-1}(\qv,\qv',\omega)=\delta(\qv-\qv')+\frac{4\pi}{q^2} \chi(\qv,\qv',\omega).
\label{ecp}
\end{equation}

For Q2D crystals, we will use the mixed
reciprocal- and real-space representation, in the $xy$ plane and $z$ direction, respectively.
Then the definition~(\ref{chi3r}) becomes
\begin{equation}
\rho_{\Gv_\|}(z,\qv_\|,\omega)= \sum\limits_{\Gv'_\|} \int\limits_{-\infty}^{\infty} \chi_{\Gv_\|\Gv'_\|}(z,z',\qv_\|,\omega) \phi^{ext}_{\Gv'_\|}(z',\qv_\|,\omega) d z',
\label{chid}
\end{equation}
where $\Gv_\|$ and $\Gv'_\|$ are the 2D reciprocal lattice vectors, and
$\qv_\|$ is the 2D wave-vector in the first Brillouin zone. 
We will also need the expression for the induced potential 
\begin{equation}
\phi^{ind}_{\Gv_\|}(z,\qv_\|,\omega) = \frac{2\pi}{|\Gv_\|+\qv_\||} \int\limits_{-\infty}^{\infty} \rho_{\Gv_\|}(z',\qv_\|,\omega) e^{-|\Gv_\|+\qv_\|||z-z'|} d z'.
\label{vind}
\end{equation}

We distinguish between different kinds of experimental setups, where it will be shown that the measurable
quantities  correspond to different definitions of the energy-loss function.

\subsection{Electron energy-loss spectroscopy}
The principal experimental method of probing $\qv_\|$- and $\omega$-dependent response of Q2D systems
is the Electron Energy-Loss Spectroscopy (EELS). We will consider separately the 
reflection and transmission EELS.

\subsubsection{Reflection EELS.}
\label{REELS}
In reflection EELS, the quantity, which
characterizes the inelastic electron scattering, is the so-called $g$-function 
\cite{Liebsch}\footnote{In this paper, we restrict ourselves to the dipole scattering regime \cite{Liebsch}.
Impact-scattering regime is more involved, requiring the solution of the inelastic scattering
problem within the distorted-waves method \cite{Nazarov-95S,Nazarov-99,Nazarov-01}, which
is outside the scope of this paper.}.
Its definition reads: Let the external potential be%
\footnote{Everywhere below, only the $z$ dependence of quantities is written down explicitly.
The implied $\rv_\|$ and $t$ dependence is always $e^{i(\qv_\|\cdot\rv_\|-\omega t)}$.}
\begin{equation}
\phi^{ext}(z,\qv_\|,\omega)= e^{q z}.
\end{equation}
Then asymptotically at $z\rightarrow +\infty$
\begin{equation}
\phi^{ind}_{\0v}(z,\qv_\|,\omega) = g(\qv_\|,\omega) e^{-q z}.
\label{vextEEPL}
\end{equation}
With the use of Eqs.~(\ref{chid}) and (\ref{vind}), the $g$-function can be written as
\cite{Persson-85,Tsuei-91,Liebsch}
\begin{equation}
g(\qv_\|,\omega)=  \frac{2\pi}{q_\|} \int\limits_{-\infty}^{\infty} e^{q_\|(z+z')} \chi_{\0v \0v}(z,z',\qv_\|,\omega) d z d z'.
\label{gchi}
\end{equation}
When $\qv_\|$ and $\omega$ are substituted with
the incident electron's loss of the parallel momentum and energy, respectively,
the differential cross-section of the inelastic electron scattering in the reflection geometry is determined by 
the energy-loss function
$-{\rm Im}\,g(\qv_\|,\omega)$, i.e., 
\begin{equation}
\frac{d^2 \sigma_{in}(\pv' \leftarrow \pv)}{d \Omega d\omega}\sim
L(\qv_\|,\omega)= -{\rm Im}\,g(\qv_\|,\omega),
\label{crsecr}
\end{equation}
where $d\Omega$ is the element of the solid-angle of the detection of the scattered electron, 
$\qv=\pv-\pv'$, $\omega=E-E'$, $E=p^2/2$, $E'=p'^2/2$, $\pv$ and $\pv'$ are the momenta of the incident and scattered electron, respectively. 
Equation~(\ref{gchi}) expresses $g$ through the density-response function $\chi$.

\subsubsection{Transmission EELS.}
\label{TEELS}
For the transmission EELS of an arbitrary (possibly, non-periodic) target,
the differential cross-section of the inelastic electron scattering is proportional to the energy-loss function
(see, e.g., \cite{Sturm-82})
\begin{equation}
\frac{d^2 \sigma_{in}(\pv' \leftarrow \pv)}{d \Omega d\omega}\sim
L(\qv_\|,\omega)= - \frac{1}{q^2} {\rm Im}\, \varepsilon^{-1}(\qv,\qv,\omega).
\label{crsec}
\end{equation}
We emphasize, that in the right-hand side of Eq.~(\ref{crsec})
$\varepsilon^{-1}$ of Eq.~(\ref{ecp}) enters {\it with $\qv$ and $\qv'$ both substituted with the same momentum
transfer $\qv$}.

We consider the case of  normal incidence. Then
\begin{equation}
\frac{(p-q_z)^2}{2}+ \frac{q_\|^2}{2}=E-\omega,
\end{equation}
\begin{equation}
q_z=p-\sqrt{p^2-2\omega-q_\|^2},
\end{equation}
and, since $p$ is large,
\begin{equation}
q_z=\frac{2\omega+q_\|^2}{2 p}.
\end{equation}

Then, by Eqs.~(\ref{crsec}) and (\ref{ecp}), and restricting ourselves to the
momentum transfer $\qv_\|$ within the first Brillouin zone, we have
\begin{equation}
L(\qv,\omega)=  - \frac{4\pi}{q^4} \, {\rm Im}  \! \int\limits_{-\infty}^{\infty} e^{i q_z (z'-z)} \chi_{\0v \0v}(z,z',\qv_\|,\omega) d z d z'.
\label{Ltransmission}
\end{equation}

Equation (\ref{Ltransmission}) is the valid energy-loss function to be used in the theory of the transmission
EELS. We point out that it replaces $- \frac{4\pi}{q^2} \, {\rm Im} \frac{1}{\epsilon(\qv,\omega)}$
\cite{Wachsmuth-13}, the latter not being defined for Q2D crystal.

\subsubsection{In-plane electronic transport.}
\label{inplane}
Let us calculate the conductivity and the energy dissipation in a Q2D crystal under the action
of the uniform in $z$ direction external
potential
\begin{equation}
\phi^{ext}(z,\qv_\|,\omega)= 1.
\label{vexttr}
\end{equation}
Then, by Eq.~(\ref{chid}),
\begin{equation}
\rho_{\0v}(z,\qv_\|,\omega)=  \int\limits_{-\infty}^{\infty} \chi_{\0v \0v}(z,z',\qv_\|,\omega)  d z',
\label{chidtr}
\end{equation}
and for the 2D particle-density averaged in the $xy$ plane
\begin{equation}
\rho_{2D}(\qv_\|,\omega)=  \int\limits_{-\infty}^{\infty} \rho_{\0v}(z,\qv_\|,\omega) dz =\int\limits_{-\infty}^{\infty} \chi_{\0v \0v}(z,z',\qv_\|,\omega)  d z d z'.
\label{chidtr2}
\end{equation}
Using the continuity equation
\begin{equation}
 \qv_\|\cdot \jv_{2D}(\qv_\|,\omega)=\omega \rho_{2D}(\qv_\|,\omega),
\end{equation}
where $\jv_{2D}$ is the 2D current-density, we can write
\begin{equation}
\jv_{2D}(\qv_\|,\omega)= \frac{ \omega \qv_\|}{q^2} \int\limits_{-\infty}^{\infty} \chi_{\0v \0v}(z,z',\qv_\|,\omega)  d z d z',
\end{equation}
and since, by Eq.~(\ref{vexttr}), $\Ev^{ext}=-i \qv_\|$,
\begin{equation}
\jv_{2D}(\qv_\|,\omega)=  \sigma^{ext}_{2D}(\qv_\|,\omega) \Ev^{ext}(\qv_\|,\omega),
\label{chidtr22}
\end{equation}
where $\sigma^{ext}_{2D}$ is the 2D conductivity {\it with respect to the external field}, which is given by
\begin{equation}
\sigma^{ext}_{2D}(\qv_\|,\omega)=  \frac{ i \omega }{q_\|^2} \int\limits_{-\infty}^{\infty} \chi_{\0v \0v}(z,z',\qv_\|,\omega)  d z d z'.
\label{chidtr222}
\end{equation}

For the energy dissipation $Q$  per unit time per unit cell area 
under the action of the external potential (\ref{vexttr})
we can write%
\footnote{Because $Q$ is the quadratic rather than linear property, we must temporarily return to the $(\rv,t)$ representation.}
\begin{equation}
Q(\qv_\|,\omega)=\frac{\omega}{2\pi A} \int\limits_0^{2\pi/\omega} d t \int\limits_{-\infty}^{\infty} d z \int\limits_A d \rv_\| \, \jv(\rv,t)\cdot \Ev(\rv,t),
\label{Q1}
\end{equation}
where $\jv(\rv,t)$ and $\Ev(\rv,t)$ are well defined 3D current-density and electric field, respectively, and  $A$ is the 2D unit cell of the Q2D crystal. Since carriers do not commit work on themselves, Eq.~(\ref{Q1}) can be rewritten as 
\begin{equation}
Q(\qv_\|,\omega)=\frac{\omega}{2\pi A} \int\limits_0^{2\pi/\omega} d t \int\limits_{-\infty}^{\infty} d z \int\limits_A d \rv_\| \, \jv(\rv,t)\cdot \Ev^{ext}(\rv,t)
\label{Q2}
\end{equation}
and simplified to
\begin{eqnarray}
Q(\qv_\|,\omega) & = \frac{1}{2} \, {\rm Re} \left[ \Ev^{ext }(\qv_\|,\omega)^* \cdot \int\limits_{-\infty}^{\infty} \jv_{\0v}(z,\qv_\|,\omega) d z \right]
\nonumber \\
&= \frac{\omega}{2 q^2} \, {\rm Re} \left[ \Ev^{ext }(\qv_\|,\omega)^* \cdot \qv_\| \int\limits_{-\infty}^{\infty} \rho_{\0v}(z,\qv_\|,\omega) d z \right].
\label{Q3}
\end{eqnarray}
Using Eqs.~(\ref{chid}) and (\ref{vexttr}), we can rewrite Eq.~(\ref{Q3}) as
\begin{equation}
Q(\qv_\|,\omega) = \frac{q^2}{2} {\rm Re} \, \sigma^{ext} (\qv_\|,\omega)=
-\frac{\omega}{2} \, {\rm Im}  \! \int\limits_{-\infty}^{\infty} 
\chi_{\0v \0v}(z,z',\qv_\|,\omega)  d z d z' .
\label{Q4}
\end{equation}

The important conclusion of Sections \ref{TEELS} - \ref{inplane} is that
the energy losses in the reflection EELS, transmission EELS,  and the energy dissipation in the in-plane transport
are governed by the quantities 
$ \! \! \int e^{q_\|(z+z')} \chi_{\0v \0v}(z,z',\qv_\|,\omega) d z d z'$,
$ \! \! \int e^{i q_z(z'-z)} \chi_{\0v \0v}(z,z',\qv_\|,\omega) d z d z'$,
and \
$ \! \! \int \chi_{\0v \0v}(z,z',\qv_\|,\omega) d z d z'$, respectively, which,
for a Q2D crystal, are clearly {\it different quantities}.
We note that in the limiting case of the purely 2D (zero-thickness) crystal, $\chi(z,z',\qv_\|,\omega)$ contains $\delta(z) \delta(z')$, and, consequently, {\it all these three quantities coincide}.

\subsubsection{Dielectric function of Q2D crystal.}
It has already been noted in Sec.~\ref{Intr} that the dielectric function cannot be rigorously
introduced for a Q2D crystal in a simple multiplicative way via Eq.~(\ref{def}). 
This is, in the first place, due to the uncertainty at which $z$ $\phi^{tot}(z,\qv_\|,\omega)$
must be used.
It is, however, convenient to keep the notion
of the dielectric function on the intuitive level, resorting to an analogy with the purely 2D system. In the latter case
\begin{equation}
\chi_{\0v \0v}(z,z',\qv_\|,\omega)= \delta(z) \delta(z') \chi^{2D}_{\0v \0v}(\qv_\|,\omega),
\label{2Dchi}
\end{equation}
where $\chi^{2D}_{\Gv_\| \Gv'_\|}(\qv_\|,\omega)$ is the 2D density-response function and the relation holds
\begin{equation}
\frac{1}{\epsilon_{2D}(\qv_\|,\omega)}= 1+\frac{2\pi}{q_\|} \chi^{2D}_{\0v \0v}(\qv_\|,\omega),
\label{e1}
\end{equation}
where $\epsilon_{2D}(\qv,\omega)$ is the dielectric function of the 2D system.
By Eqs.~(\ref{chidtr222}) and (\ref{2Dchi})
\begin{equation}
\sigma^{ext}_{2D}(\qv_\|,\omega)=\frac{i\omega}{q_\|^2} \chi^{2D}_{\0v \0v}(\qv_\|,\omega)
\end{equation}
and, using Eq.~(\ref{e1})
\begin{equation}
\frac{1}{\epsilon_{2D}(\qv_\|,\omega)}= 1+\frac{2\pi q_\|}{i \omega} \sigma^{ext}_{2D}(\qv_\|,\omega).
\label{anal}
\end{equation} 

For a Q2D crystal, {\it we define the dielectric function using} Eq.~(\ref{anal}),
where in the right-hand side we substitute $\sigma^{ext}_{2D}$ of Eq.~(\ref{chidtr222}), which is a well defined quantity
for the Q2D crystal. Therefore
\begin{equation}
\frac{1}{\epsilon_{Q2D}(\qv_\|,\omega)} \equiv 1+  \frac{2\pi}{q_\|} \int\limits_{-\infty}^{\infty}  \chi_{\0v \0v}(z,z',\qv_\|,\omega) d z d z'.
\label{epsdef}
\end{equation}

By comparing Eqs.~(\ref{Ltransmission}), (\ref{gchi}), and (\ref{epsdef}), we see that the inelastic scattering of electrons
in EELS of Q2D system is not defined by the loss-function $-{\rm Im}\, \frac{1}{\epsilon_{Q2D}(\qv_\|,\omega)}$.
It will be shown that this difference is of a quantitative importance in the interpretation
of the measurements on Q2D crystals.

Below we will see that it is convenient to have our results in the full reciprocal-space representation.
Using the Fourier expansion in the $\left[-\frac{d}{2},\frac{d}{2}\right]$ interval,
where at this point we consider $d$ an arbitrary sufficiently large distance, we can write
\begin{equation}
e^{k z}= \sum\limits_{G_{z}} b_{G_{z}}(k) e^{i G_z z},
\end{equation}
where
\begin{equation}
b_{G_z}(k) = \frac{2 \sinh\left[\frac{(k-i G_z)d}{2}\right]}{d (k- i G_z)}.
\label{bg}
\end{equation}
Then Eq.~(\ref{Ltransmission}) can be written as
\begin{equation}
L(\qv,\omega)=  -\frac{4\pi d}{q^4} {\rm Im} \! \sum\limits_{G_{z} G'_{z}} b^*_{G_z}(i q_z) b_{G'_z}(i q _z) \chi_{\0v G_z, \0v G'_z}(\qv_\|,\omega),
\label{Lt}
\end{equation}
Eq.~(\ref{gchi}) as
\begin{equation}
g(\qv_\|,\omega)=  \frac{2\pi d}{q_\|} \sum\limits_{G_{z} G'_{z}} b^*_{G_z}(q_\|) b_{G'_z}(q_\|) \chi_{\0v G_z, \0v G'_z}(\qv_\|,\omega),
\label{gbg}
\end{equation}
and Eq.~(\ref{epsdef}) as
\begin{equation}
\frac{1}{\epsilon_{Q2D}(\qv_\|,\omega)}= 1+  \frac{2\pi d}{q_\|}  \chi_{\0v 0, \0v 0}(\qv_\|,\omega).
\label{epsfullrec}
\end{equation}

Having established  expressions for the observables via the density-response function
of a single layer
$\chi_{\Gv_\| \Gv'_\|}(z,z',\qv_\|,\omega)$, we  turn to the methods of the calculation of the latter.

\section{Excitation of  Q2D system from the super-cell geometry calculation}
\label{Extr}

\begin{figure}[h!]
\begin{center}
\includegraphics[width= 0.5 \columnwidth, trim= 0 0 0 0, clip=true]{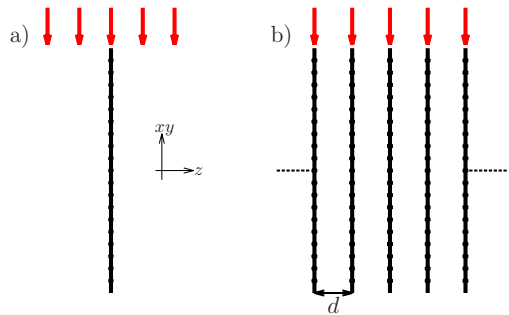} 
\end{center}
\caption{\label{arr} Schematics of  2D material under an external field.
a) 2D single-layer geometry and b) 3D super-cell geometry.}
\end{figure}

By far the widest used at this time way of the calculation of the electronic response of 
atomically thin Q2D crystals is the super-cell geometry approach.
This is, instead of considering a single layer [schematized in Fig.~\ref{arr}, a)],
to artificially periodically replicate it in $z$-direction [Fig.~\ref{arr}, b)],
and study the resulting 3D periodic system in place of the original single-layer one.
It is, however, known \cite{Rozzi-06,Despoja-13,Nazarov-14} that the  3D dielectric
function obtained in the super-cell geometry has nothing to do with that of a single-layer system of interest, unless a special procedure
of extracting the 2D response from the 3D super-cell geometry calculation (i.e., the elimination
of the spurious inter-layer interaction) is applied.
In \ref{AExtr} we give a detailed derivation of the expression for the 
density-response function $\chi$ of the single-layer system from
$\tilde{\chi}$ - the density-response function of the array system.
It reads in the matrix form%
\footnote{In Eq.~(\ref{fn_text}), the full 3D reciprocal space representation for both $\tilde{\chi}$
and $\chi$ is used, which is natural for the former, but may look artificial for the latter.
We, however, note that an arbitrary function of $z$ and $z'$ can be expanded in the Fourier series in 
$z,z' \in [-\frac{d}{2},\frac{d}{2}]$, and then, if necessary, the real-space representation  can be retrieved
by the inverse Fourier transform.}
\begin{equation}
\chi(\qv_\|,\omega)= \tilde{\chi}(\qv_\|,\omega) \left[1+ C(\qv_\|) \tilde{\chi}(\qv_\|,\omega) \right]^{-1},
\label{fn_text}
\end{equation}
where the matrix $C_{\Gv \Gv'}$ is given by 
\begin{eqnarray}
C_{\Gv  \Gv' }(\qv_\|) &=F_{G_z G_z'}(|\Gv_\|+\qv_\||) \delta_{\Gv_\| \Gv'_\|}, \\
F_{G_z G_z'}(p)&  =\frac{4\pi (p^2-G_z G_z')}{p d (p^2+G_z^2)(p^2+{G'}_z^2)} \cos \left[ \frac{(G_z+G_z') d}{2} \right]
 (1-e^{-p d}) .
\label{fn_text_F}
\end{eqnarray}

The density-response function of the array system $\tilde{\chi}_{\Gv_\| \Gv'_\|}(\qv_\|,\omega)$ can be routinely obtained with
the 3D solid-state periodic codes such, e.g., as Elk, abinit, Wien2k, {\it etc.}.
While $\tilde{\chi}$ is different for different layers' separations $d$, 
$\chi$ obtained through Eq.~(\ref{fn_text})
{\em does not depend on} $d$, 
which is crucial for its being a true characteristic of a stand-alone 2D system%
\footnote{Strictly speaking, this is $\chi_{\Gv_\| \Gv'_\|}(z,z',\qv_\|,\omega)$ which is independent on $d$.
Its Fourier transform in the interval $z,z' \in [-\frac{d}{2},\frac{d}{2}]$ does formally depend on $d$, which is of no physical consequence.}. We note that Eqs.~(\ref{fn_text})- (\ref{fn_text_F}) are valid
not only in RPA \cite{Despoja-13}, but also in TDDFT with (semi-)local $f_{xc}$ (such as, e.g., ALDA), when 
different layers interact by the Coulomb potential only.

We conclude this section by noting that the ignoring the fundamental difference between
the single-layer response of a Q2D crystal and that of the array of those layers has led to the misinterpretation
of the $\pi$ and $\pi+\sigma$ peaks as single-particle inter-band transitions rather than plasmons \cite{Nelson-14}.

\section{Results of calculations and discussion}
\label{RES}
\begin{figure} [h] 
\includegraphics[width= 1 \columnwidth, trim= 20 0 10 0, clip=true]{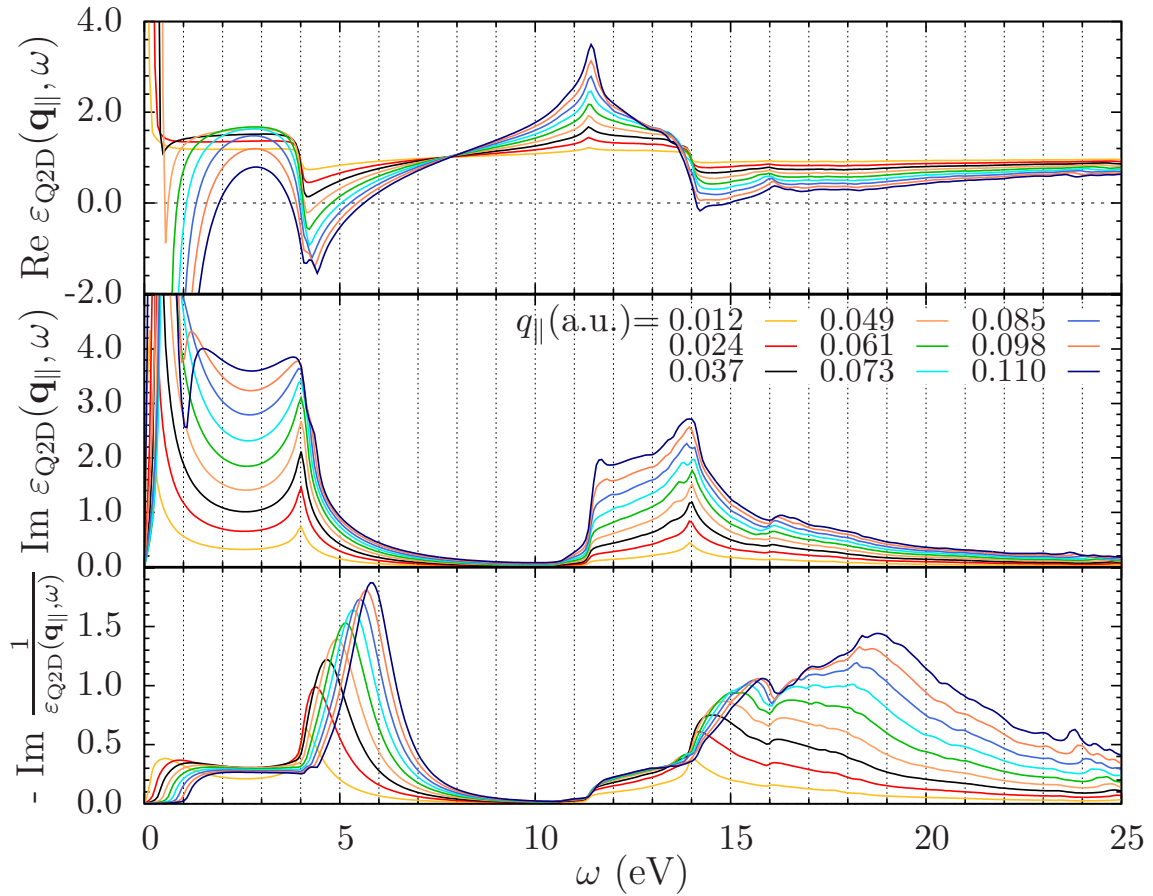} 
\caption{\label{eps} 
Dielectric function of the pristine monolayer graphene obtained with Eq.~(\ref{epsfullrec})
and the corresponding energy-loss function.
Momentum $\qv_\|$ is varied along the $\Gamma M$ direction.
}
\end{figure}

\begin{figure} [h] 
\includegraphics[width= 1 \columnwidth, trim= 0 0 0 0, clip=true]{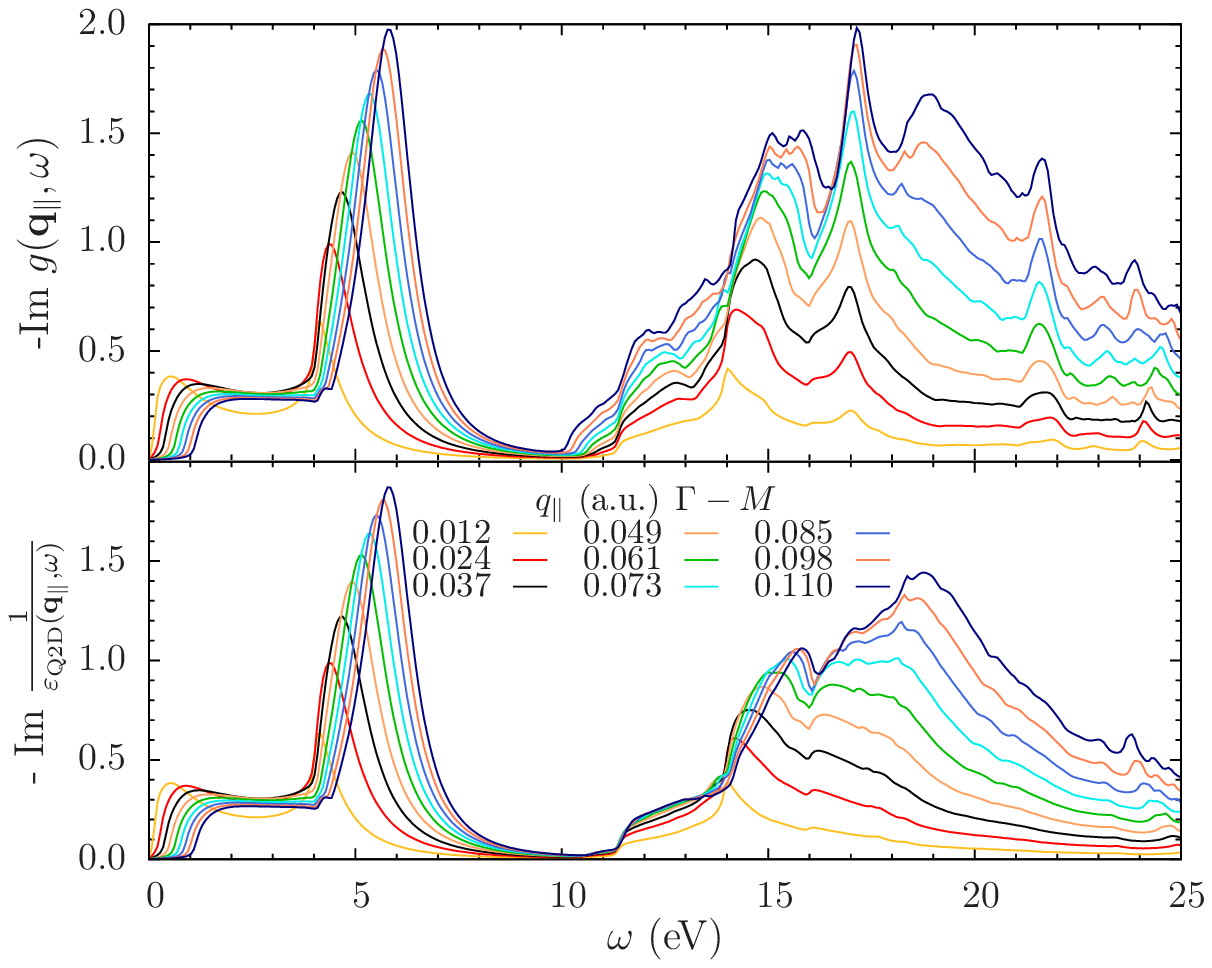} 
\caption{\label{L2g2} 
The EELS-related energy-loss function $-{\rm Im}\, g(\qv_\|,\omega)$ of Eq.~(\ref{gbg}) (upper panel) compared with the energy-loss function $-{\rm Im}\, \frac{1}{\varepsilon_{Q2D}(\qv_\|,\omega)}$ of Eq.~(\ref{epsfullrec})
(lower panel) for pristine monolayer graphene. 
}
\end{figure}

\begin{figure} [h] 
\includegraphics[width= 1 \columnwidth, trim= 30 0 0 0, clip=true]{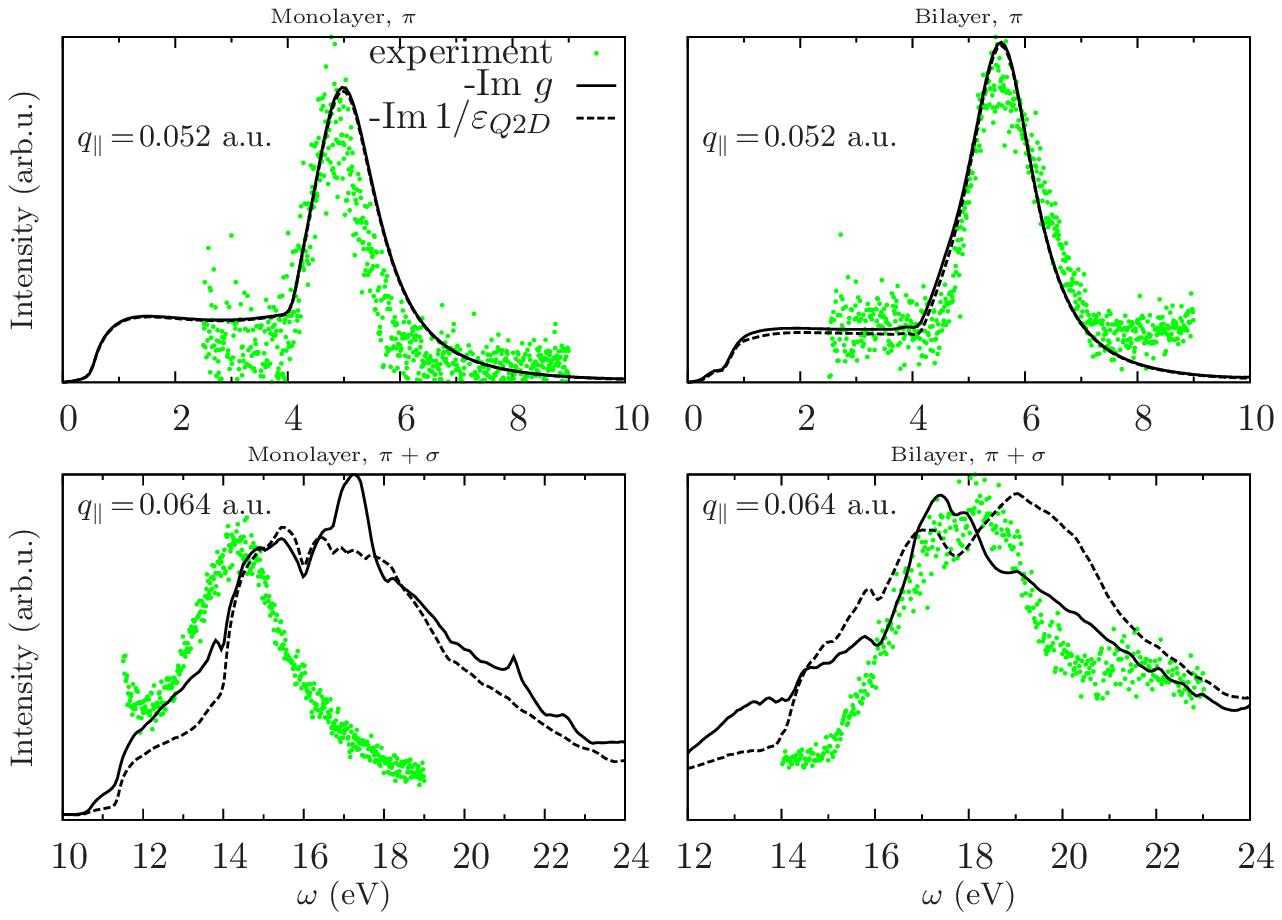}
\caption{\label{explu} 
The reflection EELS energy-loss function $-{\rm Im}\, g(\qv_\|,\omega)$ of Eq.~(\ref{gbg})
(solid lines) and the energy-loss function $-{\rm Im}\, \frac{1}{\varepsilon_{Q2D}(\qv_\|,\omega)}$ of Eq.~(\ref{epsfullrec}) (dashed lines) for monolayer (left) and bilayer (right)  graphene
in the energy-range of $\pi$ (upper) and $\pi+\sigma$ (lower) plasmons.
Symbols are experimental reflection EELS of Ref.~\cite{Lu-09}.
The theoretical spectra have been roughly normalized to the experimental ones.}
\end{figure}

\begin{figure} [h] 
\includegraphics[width= 1 \columnwidth, trim= 0 0 0 0, clip=true]{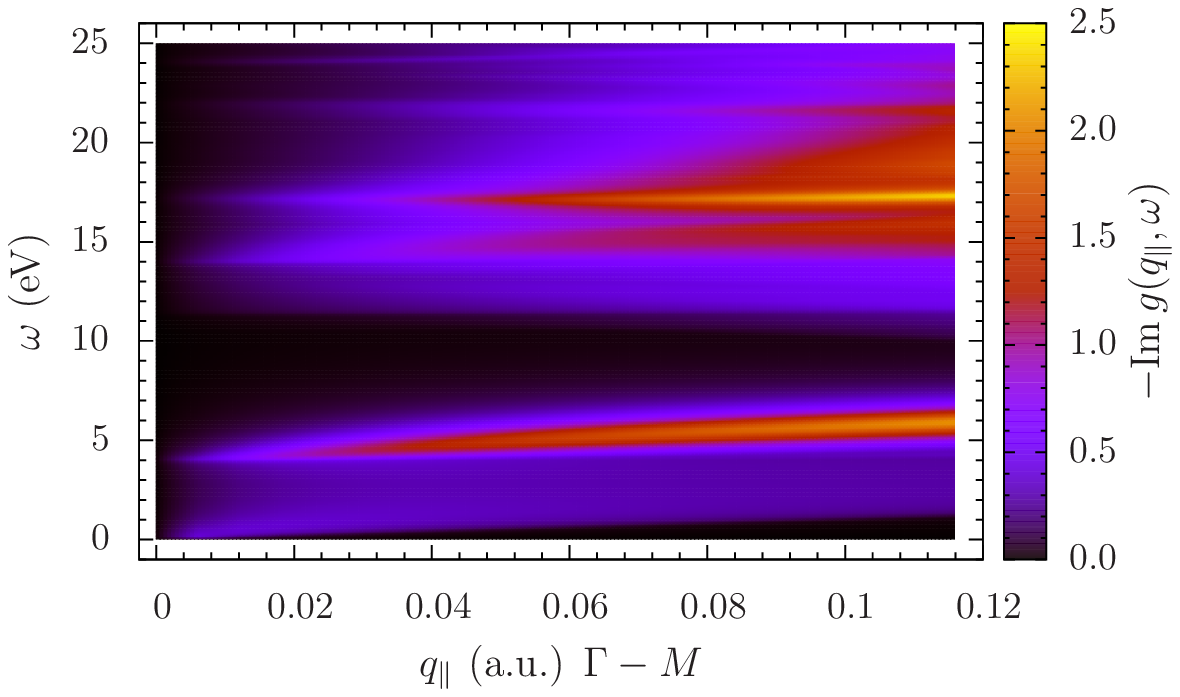} 
\caption{\label{disp} 
Dispersion of $\pi$ and $\pi+\sigma$ plasmons.
Momentum $\qv_\|$ is changed along the $\Gamma M$ direction.
}
\end{figure}

We have conducted  the super-cell geometry calculation for the monolayer pristine graphene {\em followed by the application
of Eq.~(\ref{fn_text})} for the extraction of the single-layer response 
(elimination of the spurious inter-layer interaction) from the 3D calculation. In the DFT super-cell calculation, 
we used  the full-potential linear augmented plane-wave (FP-LAPW) code Elk \cite{Elk}. 
The $z$-axis period of the super-cell was taken 20 a.u.
The $k$-point grid of $512\times 512 \times 1$, 30 empty bands, and  the damping parameter of 
0.002 a.u.
were used in both the ground-state and the linear-response calculations. 
The former was conducted within the local-density approximation for the exchange-correlation potential, 
while the latter was  the random-phase approximation (RPA) one (i.e., the exchange-correlation kernel $f_{xc}$ was set to zero).
The dielectric matrix of the 3D system was of the size $55\times 55$, which was inverted to obtain
$\varepsilon_{3D}(\qv_\|,\omega;d)$ with the local-field effects included both in the perpendicular
and in-plane directions. 

Results for the Q2D dielectric and energy-loss functions 
obtained through Eq.~(\ref{epsfullrec})
are presented in Fig.~\ref{eps}. Features at the energy-loss function around 5 and 15 eV are usually
referred to as $\pi$ and $\pi+\sigma$ peaks, respectively, \cite{Gass-08,Eberlein-08,Liu-08,Nelson-14,Liou-15} (lower panel of Fig,~\ref{eps}). To determine whether they are collective excitations (plasmons) \cite{Gass-08,Eberlein-08,Liu-08,Liou-15} or single-particle (inter-band) transitions \cite{Nelson-14}, we scrutinize  
the dielectric function. 

A clear criterion for an excitation to be classified as plasmon
is the requirement of the real-part of the dielectric function to cross zero at the corresponding
energy. Otherwise, if  the peak at the energy-loss function is due to a
peak at the imaginary part of the dielectric function, the excitation is a single-particle transition.
In our case, starting from greater wave-vectors (from $q\approx$ 0.05 and 0.1 a.u.,
for $\pi$ and $\pi+\sigma$ peaks, respectively) the real part of the dielectric function does
cross zero, and, thereby, from those $q$-s up the peaks are  plasmons unambiguously. Are they
also plasmons at smaller $q$-s, or they turn into single-particle transitions, although keeping
full similarity to their higher-$q$ plasmon counterparts ? 
To answer this question, we need to establish whether, although ${\rm Re}\,\epsilon_{2D}$ does not cross zero
at smaller wave-vectors, the features at the energy-loss function are due to ${\rm Re}\,\epsilon_{2D}$ {\em approaching} zero, or they are due to peaks at ${\rm Im}\,\epsilon_{2D}$ (cf., \cite{Nazarov-14}). 

As can be  seen in Fig.~\ref{eps}, middle panel, Im~$\varepsilon_{2D}$
has prominent {\em non-dispersive} (standing in place with the variation of $q_\|$) feature at $\omega\approx$ 4 eV. A prominent {\em dispersive} peak in the energy-loss function (bottom panel) follows the position of 
{\em not only zero} (at greater values of $q_\|$) but also the {\em  minimum} (at smaller $q_\|$-s)
of Re~$\varepsilon_{2D}$ in the range of 4 - 6 eV, and must, therefore, in both cases be classified  
as $\pi$ plasmon.

The situation is similar, though less transparent, in the case of the $\pi+\sigma$ peak.
Here, the broad feature at $L(\qv_\|,\omega)$,
which spreads from $\omega\approx$ 11 eV upward, is due to the overlap of the single-particle
and collective excitations. Up to $\omega\approx$ 14 eV, the excitation is purely due to the single-particle
transitions and is non-dispersive. Then, from 14 eV upward, a dispersive feature due
to zero or small values of ${\rm Re}\,\varepsilon_{2D}$ begins. Because of ${\rm Re}\,\varepsilon_{2D}$
having a low slope and remaining small in absolute value in a wide $\omega$ range, this feature becomes very broad. 
It, however, is clearly plasmon as well.

Interestingly, in two points, at $\omega\approx$ 8 and 14 eV, 
${\rm Re}\,\varepsilon_{2D}$ becomes nearly unity for all values of $q_\|$, so that all 
the curves in  Fig.~\ref{eps}, upper panel, intersect at these two points. The physical reason
for this is still to be understood.

In Fig.~\ref{L2g2}, the reflection EELS-related energy-loss function $-{\rm Im}\,g(\qv_\|,\omega)$, calculated by Eq.~(\ref{gbg}), 
is compared with the loss-function $-{\rm Im}\,\frac{1}{\varepsilon_{Q2D}(\qv_\|,\omega)}$ of Eq.~(\ref{epsfullrec}), the latter relevant to the energy dissipation in the in-plane transport .
While the $\pi$ features ($\sim$ 5 eV) are similar for both loss-functions,
the $\pi+\sigma$ features are considerably different, the EELS loss-function having a richer structure in this energy range. This can be understood by noting that, according to Eq.~(\ref{epsfullrec}), $1/\varepsilon$ is determined
by the head element of the density-response matrix $\chi_{\0v 0 \0v 0}(\qv_\|,\omega)$, while, by Eq.~(\ref{gbg}), the $g$-function also includes contributions from $\chi_{\0v G_z \0v G'_z}(\qv_\|,\omega)$ with nonzero $G_z$ and/or $G'_z$.
The role of the latter increases with the increase of $\omega$,
transitions into the higher bands becoming allowed, which leads to the differences in the spectra.

In Fig.~\ref{explu} we compare our theory with reflection EELS experiment
of Ref.~\cite{Lu-09} on monolayer and bilayer graphene. 
In the energy range of $\pi$ plasmon (two upper graphs in Fig.~\ref{explu}), $g$-function and 
the energy-loss ($L$) function are practically identical and both compare to experiment rather well.
On the contrary, in the range of $\pi+\sigma$ plasmons, the $g$- and $L$-functions
acquire  differences for both monolayer and, particularly, for bilayer graphene (lower graphs).
For bilayer graphene $\pi+\sigma$ peak  at $g$-function is narrower than at the $L$-function,
the former being closer to experiment (lower right).  Remarkably, for monolayer graphene both $g$- and $L$-functions find themselves in shear disagreement with experiment (lower left).

Obviously, the main source  of the disagreement between the theory and experiment can be the presence of the
SiC (0001) substrate in the latter and the absence thereof in the former. The substrate may cause additional screening leading to the shift of plasmon peaks and change the dispersion law \cite{Politano-11,Politano-12,Generalov-12,Cupolillo-15}.
Moreover, the interaction of graphene with a substrate can cause the deformation of the graphene sheet,
leading to the confinement of the plasmon oscillations between the ripples \cite{Politano-2013-2}. 

Let us briefly discuss the modifications to the theory necessary to include Q2D crystals supported on substrates.
For the {\it ab initio} theoretical treatment of surfaces, either pure or adsorbates covered, the super-cell method is widely applied as well. 
In variance with the stand-alone Q2D systems, for surfaces, rather thick slabs of the material alternated with the similar thickness vacuum slabs are used. Equations (\ref{fn_text})-(\ref{fn_text_F})
are readily applicable in this case too, producing as an output the density-response function of a {\it single slab}. This slab having two surfaces is not, however,  the semi-infinite system of interest. 
Neither the array of such slabs, from which we have started, is. Which system,
the array of slabs or a single slab, is better to model a Q2D crystal on top of a semi-infinite substrate must be decided in each particular case. An alternative  method may be the inclusion of the substrate
by means of the effective background dielectric function after having calculated the response of a stand-alone Q2D system. Although it may be useful in some situations, this would necessarily be a phenomenological approach, probably inconsistent with the rigorousness of the calculation
for the Q2D crystal alone. Finally, the ultimate theory will be that asymptotically matching Q2D crystal
on the surface with the 3D periodic bulk of the substrate, which is a very challenging task.

The neglect of the many-body exchange-correlation effects \cite{Botti-04,Nazarov-11},
both static and dynamic, may further contribute to the discrepancy.
Both the accurate inclusion of the substrate in the theoretical setup and the inclusion of the many-body effects
by means of using nonlocal  exchange-correlation kernel of TDDFT remain the major challenges of the present-day 
theory of Q2D materials. 

Figure~\ref{disp} shows the dispersion of $\pi$ and $\pi+\sigma$
plasmons with the $\qv_\|$ vector varied in the $\Gamma M$ direction calculated in the EELS setup. The $\pi+\sigma$ plasmon has a complex structure and is split into distinct sub-excitations.

\begin{figure} [h] 
\includegraphics[width= 1 \columnwidth, trim= 30 0 0 0, clip=true]{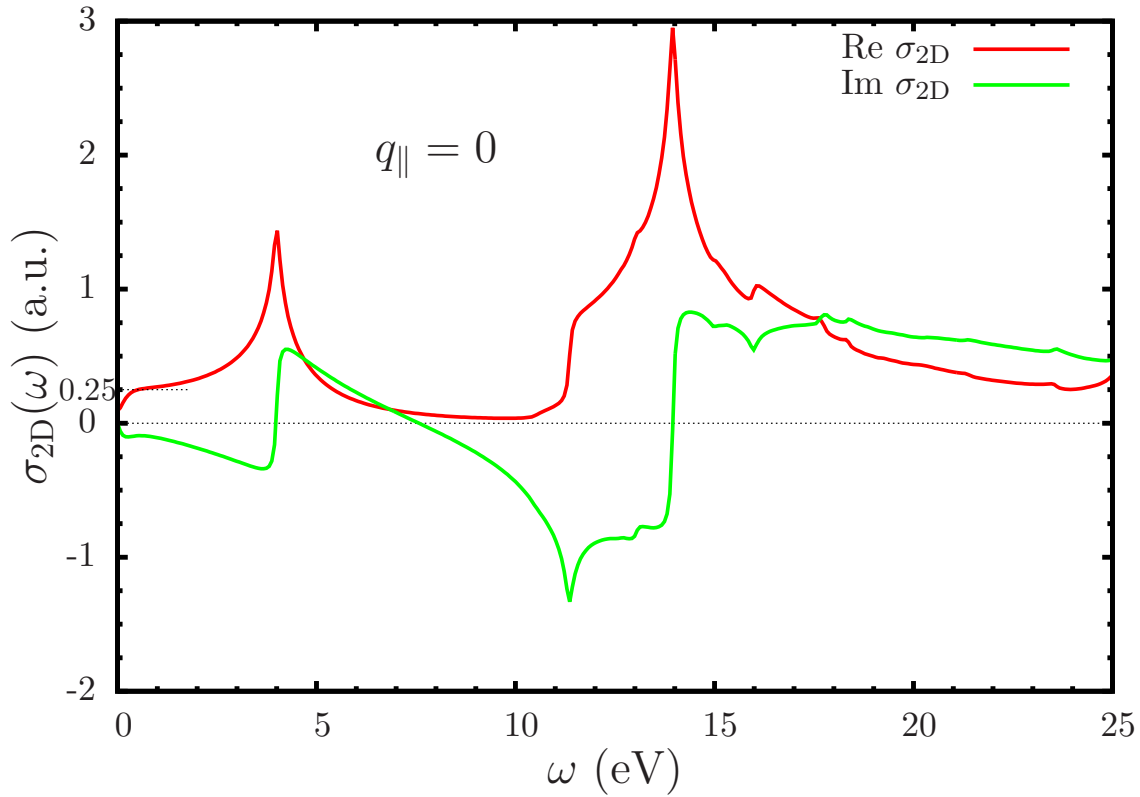} 
\caption{\label{sig} Frequency-dependent conductivity of pristine monolayer graphene at $q_\|=0$.
When $\omega\rightarrow 0$, the conductivity must converge to its analytic two-bands model value of $1/4$,
which it does, except for too small $\omega$ (at the very tip of the Dirac cone), where the accuracy of the calculation is limited by the $k$-point grid.
}
\end{figure}

As noted in Ref.~\cite{Nelson-14}, in a 2D system there can be no plasmon at $q_\|=0$.
In the ordinary 2D electron gas \cite{Stern-67}
and in doped graphene in the case of 2D plasmon due to electrons in the $\pi^*$ band \cite{Sarma-09}, this is ensured by the plasmon disappearance via its energy tending
to zero as $\sqrt{q_\|}$ when $q_\|\rightarrow 0$.
However, as can be seen from Fig.~\ref{eps}, lower panel, for $\pi$ and $\pi+\sigma$ plasmons in graphene the same realizes in a different way: 
The energy positions of these plasmons 
converge  to finite values (4 and 14 eV for $\pi$ and $\pi+\sigma$ plasmons, respectively)
as $q_\|\rightarrow 0$, while the amplitudes of the corresponding peaks go to zero.
In this regard, as in many others, graphene is also special compared to the ordinary 2D electron gas.
The zero limit of the plasmons' amplitude can be understood from simple arguments.
Indeed, quite generally, in a 2D crystal, the relation between the  dielectric function $\varepsilon_{2D}$
and the  conductivity $\sigma_{2D}$ is
\begin{equation}
\varepsilon_{2D}(\qv_\|,\omega)= 1+\frac{2\pi i q_\| \sigma_{2D}(\qv_\|,\omega)}{\omega}.
\end{equation}
Since in the single-layer graphene, $\sigma_{2D}(\qv_\|=\0v,\omega)$ is finite [and known \cite{Wunsch-06} to 
tend to $e^2/4 \hbar $ as $\omega\rightarrow 0$],
$\varepsilon_{2D}(\qv_\|=\0v,\omega)$ is unity identically, leading to the zero energy-loss function.
Our calculated conductivity of the monolayer pristine graphene at $q_\|=0$ versus the frequency
is shown in Fig.~\ref{sig}, where the static limit of $1/4$ is shown by the dotted line.

\section{Conclusions}
\label{Concl}
We have identified and dealt with the extra challenges the {\it ab initio} theory of electronic excitations in 
quasi-2D crystals presents in  comparison to both the bulk and purely 2D cases.
In particular, we have re-examined the problem of the correspondence  between the theoretically
calculated quantities and the observables  in the measurements on quasi-2D crystals and 
found that the energy-loss function $- {\rm Im}\,\frac{1}{\epsilon(\qv_\|,\omega)}$,
conventionally used for the interpretation of the EELS data, is not, generally speaking, the right quantity
to be compared with this kind of the experiment. Instead, the EELS-related energy-loss functions,
specific for both transmission and reflection experimental setups,
must be used, which has been shown to be both qualitatively and quantitatively different in the case of quasi-2D systems. In the limit of the zero crystal thickness, our theory reduces to the conventionally used one.

We have addressed the problem of the classification of the
$\pi$ and $\pi+\sigma$ peaks in monolayer graphene. 
By the use of the accurate procedure of extracting the dielectric function of a 2D crystal
from the 3D super-cell geometry calculation, we have shown conclusively that there exist prominent 
$\pi$ and $\pi+\sigma$ collective exitations (plasmons) in graphene, although they are accompanied 
by interband transitions in the close energy ranges.
Plasmons and single-particle interband  transitions can be  distinguished from each other
provided the wave-vector-resolved analysis is performed. 

We have also demonstrated the
importance of the correct interpretation of the super-cell geometry calculation results, the latter
taken simplistically, can lead to erroneous qualitative conclusions for  materials of reduced dimensionality.
We expect that the present theory of the electronic excitations in Q2D crystals, by  correctly accounting 
for the atomic or nano- thickness of the systems studied, will further boost the theoretical and experimental
work, improving  the methods of comparison between the two.

\ack
I am indebted to Guang-Yu Guo for extremely fruitful discussions and
I  thank authors of Ref.~\cite{Lu-09} for providing  the digital data of their experimental EELS spectra.
The support from the Ministry of Science and Technology, Taiwan,
Grant No. 103-2112-M-001-007, is acknowledged.

\appendix

\section{Extraction of the response of a single layer from that of the array-of-layers system}
\label{AExtr}
Let us consider a periodic array of identical layers, as schematized in Fig.~\ref{arr}, and let us construct
the density-response function $\tilde{\chi}_{\Gv_\| \Gv'_\|}(z,z',\qv_\|,\omega)$ of the array system. We assume that the separation $d$ between the layers is large enough so that  densities, both the ground-state and the dynamically induced, do not overlap between layers. Then for the array system we can write

\begin{equation}
n_{\Gv_\|}(z,\qv_\|,\omega)= 
\sum\limits_{\Gv'_\|}
\int\limits_{-d/2}^{d/2} \chi_{\Gv_\| \Gv'_\|}(z,z',\qv_\|,\omega) \phi^{eff}_{\Gv'_\|}(z',\qv_\|,\omega) d z',
\label{na}
\end{equation}
where
\begin{eqnarray}
\! \! \! \! \! \! \! \! \! \! \! \! \! \! \! \!
\phi^{eff}_{\Gv}(z,\qv_\|,\omega) &= \phi^{ext}_{\Gv}(z) \nonumber \\
&+
\frac{2\pi}{|\Gv+\qv_\||} \sum\limits_{m=-\infty}^{\infty} {} \! \! \! ^{'}
\int\limits_{-d/2}^{d/2} n_\Gv(z',\qv_\|,\omega)e^{-|\Gv+\qv_\|| |z-z'-m d|} d z'.
\label{sum}
\end{eqnarray}
Equation~(\ref{na})  uses the fact that the potential external to the $m=0$ layer
is that of Eq.~(\ref{sum}), i.e., it is the proper external potential plus the potentials
induced by all the layers with $m\ne 0$. The prime at the sum means that the $m=0$ term is excluded from the summation. Performing the summation explicitly, we have%
\footnote{Equation (\ref{veff}) corrects an error in Eq.~(A3) of Ref.~\cite{Nazarov-14}.
However, all the formulas derived in Ref.~\cite{Nazarov-14} under the assumption of $q a \ll 1$,
where $a$ is the effective width of a layer, remain valid. }
\begin{eqnarray}
\hspace{-2.5 cm}
\phi^{eff}_{\Gv}(z,\qv_\|,\omega) & = \phi^{ext}_{\Gv}(z,\qv_\|,\omega) \nonumber \\
&+
\frac{4\pi}{|\Gv \! + \! \qv_\||(e^{|\Gv+\qv_\|| d}  \! - \! 1)} \!
\int\limits_{-d/2}^{d/2} \! \! \! n_\Gv(z',\qv_\|,\omega) \cosh [|\Gv+\qv_\|| (z-z')] d z'.
\label{veff}
\end{eqnarray}
Expanding into the Fourier series
\begin{equation}
\cosh [p(z-z')]  = \sum\limits_{G_z G'_z} D_{G_z G'_z}(p) e^{i G_z z -i G'_z z'},
\end{equation}
where
\begin{equation}
D_{G_z G'_z}(p)=   \frac{4(p^2-G_z G'_z)}{d^2 (p^2+G_z^2)(p^2+{G'_z}^2)} \cos \left[ \frac{(G_z+G'_z) d}{2} \right]
 \sinh^2 \left( \frac{p d}{2} \right) ,
\label{DD}
\end{equation}
and introducing the matrices
\begin{eqnarray}
F_{G_z G'_z}(p)& = \frac{4\pi d}{p (e^{p d}-1)} D_{G_z G'_z}(p)
 \nonumber \\
 & =\frac{4\pi (p^2-G_z G'_z)}{p d (p^2+G_z^2)(p^2+{G'_z}^2)} \cos \left[ \frac{(G_z+G'_z) d}{2} \right]
 (1-e^{-p d}) 
\end{eqnarray}
\begin{equation}
 C_{\Gv_\| G_z, \Gv'_\| G'_z}(\qv_\|) =F_{G_z G'_z}(|\Gv_\|+\qv_\||) \delta_{\Gv_\| \Gv'_\|},
 \label{CD}
\end{equation}
we arrive at the fully 3D reciprocal-space representation
\begin{equation}
\phi^{eff}_{\Gv}(\qv_\|,\omega) = \phi^{ext}_{\Gv}(\qv_\|,\omega) + \sum\limits_{\Gv'} C_{\Gv \Gv'}(\qv_\|)  n_{\Gv'}(\qv_\|,\omega),
\label{vvv}
\end{equation}
\begin{equation}
n_{\Gv }(\qv_\|,\omega)= \sum\limits_{\Gv' } \chi_{\Gv  \Gv' }(\qv_\|,\omega) \phi^{eff}_{\Gv'}(\qv_\|,\omega).
\label{nnn}
\end{equation}

Using Eqs.~(\ref{vvv}) and (\ref{nnn}), the definition of $\tilde{\chi}$ as the density-response function of the array system
\begin{equation}
n_{\Gv }(\qv_\|,\omega)= \sum\limits_{\Gv' } \tilde{\chi}_{\Gv \Gv' }(\qv_\|,\omega) \phi^{ext}_{\Gv' }(\qv_\|,\omega),
\end{equation}
and in view of the arbitrariness of $\phi^{ext}$, we can write in the matrix form
\begin{equation}
\tilde{\chi}(\qv_\|,\omega) =\chi(\qv_\|,\omega) +\chi(\qv_\|,\omega) C \tilde{\chi}(\qv_\|,\omega).
\label{but1}
\end{equation}
Finally, inverting Eq.~(\ref{but1}), we arrive at Eq.~(\ref{fn_text}).

It may be useful to consider approximations to the above exact scheme. First,
since $\Gv_\| d \gg 1$ for $\Gv_\| \ne \0v$, this is practically exact to keep in Eq.~(\ref{sum})  the $\Gv_\|=0$ term only.
Further, if $q a \ll 1$, where $a$ is the effective width of the layer, Eq.~(\ref{fn_text}) can be shown to yield
a convenient relation between the corresponding 3D and 2D dielectric functions \cite{Nazarov-14}
\begin{equation}
\frac{1}{\varepsilon_{Q2D}(\qv_\|,\omega)}=
1+\frac{1}{2} \frac{1}{ \frac{1}{\left[ \frac{1}{\varepsilon_{3D}(\qv_\|,\omega;d)} -1 \right] q_\| d} +
\frac{1}{e^{q_\| d}-1}}  ,
\label{eps2d3d}
\end{equation}
where  $\varepsilon_{3D}(\qv_\|,\omega;d)$ is the dielectric function of the fictitious  periodic
3D system comprised of the layers separated  by the distance $d$.

In the $q_\| d\gg 1$ limit we have from Eq.~(\ref{eps2d3d})
\begin{equation}
\frac{1}{\varepsilon_{2D}(\qv_\|,\omega)}=
1+\frac{q_\| d}{2} \left[ \frac{1}{\varepsilon_{3D}(\qv_\|,\omega;d)} -1 \right].
\label{eps2d3di}
\end{equation}

In the long-wave limit $q_\| d\ll 1$
\begin{equation}
\frac{1}{\varepsilon_{2D}(\qv_\|,\omega)}=
1+\frac{q_\| d}{2} \left[ 1-\varepsilon_{3D}(\qv_\|,\omega;d)\right],
\label{eps2d3d00}
\end{equation}
which, since the second term is small, can be written as
\begin{equation}
\varepsilon_{2D}(\qv_\|,\omega)=
1-\frac{q_\| d}{2} \left[ 1-\varepsilon_{3D}(\qv_\|,\omega;d)\right].
\label{eps2d3d0}
\end{equation}
\


\providecommand{\newblock}{}

\end{document}